\documentclass[prd,showpacs,floatfix,amsmath,amssymb,floatfix,12pt,english]{revtex4}
\usepackage[T1]{fontenc}
\usepackage[utf8]{inputenc}
\usepackage{babel}
\usepackage{units}
\usepackage{amsmath}
\usepackage{esint}
\usepackage[normalem]{ulem}
\usepackage[unicode=true,pdfusetitle,
 bookmarks=true,bookmarksnumbered=true,bookmarksopen=true,bookmarksopenlevel=2,
 breaklinks=false,pdfborder={0 0 1},backref=false,colorlinks=false]
 {hyperref}

\begin{document}

\title{Radial excitations of mesons and nucleons from QCD sum rules}

\author{Jin-Feng Jiang}
\email{jfjiang@pku.edu.cn} \affiliation {School of Physics and State
Key Laboratory of Nuclear Physics and Technology, Peking University,
Beijing 100871, China}

\author{Shi-Lin Zhu}
\email{zhusl@pku.edu.cn} \affiliation{
School of Physics and State Key Laboratory of Nuclear Physics and Technology, Peking University, Beijing 100871, China \\
Collaborative Innovation Center of Quantum Matter, Beijing 100871, China \\
Center of High Energy Physics, Peking University, Beijing 100871,
China}

\begin{abstract}

\pacs{12.38.Lg, 14.40.-n, 14.20.Dh}

Within the framework QCD sum rules, we use the least square fitting
method to investigate the first radial excitations of the nucleon
and light mesons such as $\rho$, $K^{*}$, $\pi$ , $\varphi$. The
extracted masses of these radial excitations are consistent with the
experimental data. Especially we find that the decay constant of
$\pi(1300)$, which is the the first radial excitation of $\pi$, is
tiny and strongly suppressed as a consequence of chiral symmetry.

\end{abstract}

\maketitle

\section{INTRODUCTION}
\label{sec2}

The method of QCD sum rules  has been widely used to extract the
resonance information in hadron physics \cite{Shifman1979}. This
formalism is usually applied to study the ground state in a specific
channel due to the limitation of theoretical accuracy and difficulty
of numerical analysis. The excitations of mesons have been studied within finite energy sum rules in the literature \cite{Krasnikov1982,Kataev1983,Gorishnii1984}. Recently, there are some attempts to study
the excitations of the heavy-light mesons using the method of QCD
sum rules \cite{Gelhausen2014}.

The radial excitations have the same spin-parity as the ground
state. Experimentally many radial excitations of mesons and baryons
have been established \cite{Olive2014}. Sometimes it is quite
difficult to identify the radial excitations of hadrons. For
example, the situation of the radial excitations of the vector
charmonium above 4 GeV becomes quite unclear after so many
charmonium-like XYZ states have been reported experimentally in the
past decade. Theoretical investigations of the radial excitations
are also very challenging.

In this work, we shall study the first radial excitations of the
light mesons and nucleon within the framework of the QCD sum rule
formalism. We explicitly keep two poles in the usual spectrum
representation. Then we employ the least square method in the
numerical analysis to extract the resonance information of the first
radial excited state. The extracted masses of the radial excitations
of the light mesons and nucleon agree with the experimental data
quite well.

The paper is organized as follows. In Section \ref{sec2}, we
introduce the QCD sum rule formalism and our least-square
method. The numerical results are presented in Sections
\ref{sec3rho}-\ref{sec3nucleon}. The last section is a short
summary.

\section{Formalism}
\label{sec2}

Within the framework of the QCD sum rule approach, we study the
correlation function at the quark level
\begin{equation}
\Pi\left(q\right)={\rm
i}\int\mathrm{d}^{4}x\,\mathrm{e}^{\mathrm{i}qx}\left\langle
0\left|T\left\{ j\left(x\right)j^{\dagger}\left(0\right)\right\}
\right|0\right\rangle
\end{equation}
where $j\left(x\right)$ is the interpolating current with the same
quantum numbers as the hadrons. The above correlation function
satisfies the dispersion relation
\begin{equation}
\Pi(q^{2})=\frac{1}{\pi}\int_{s_{min}}ds\frac{\mathrm{Im}\Pi(s)}{s-q^{2}-\mathrm{i}\epsilon},
\end{equation}
At the quark gluon level, the correlation function can be calculated
with the operator product expansion. The gluon and quark condensates
appear as higher dimensional operators in this expansion. At the
hadron level, the spectral density of the correlation function can
be expressed in terms of the hadron masses and couplings. Due to the
quark hadron duality, we get an equation called the QCD sum rule
which relates the correlation function at the quark gluon level to
the physical states. After making Borel transformation to the sum
rule in the momentum space, one gets
\begin{equation}
\Pi^{\prime}\left(M^{2}\right)=\frac{1}{\pi}\int\mathrm{e}^{-\nicefrac{s}{M^{2}}}\mathrm{Im}\Pi\left(s\right){\rm
ds}
\end{equation}
where $M$ is the Borel parameter.

The spectral density usually takes the one-pole approximation
\begin{equation}
\rho(s)\equiv\frac{1}{\pi}\mathrm{Im}\Pi\left(s\right)=f\delta\left(s-m^{2}\right)+\rho_{continuum}\theta\left(s-s_{0}\right),\label{eq:usual
spectrum}
\end{equation}
where $m$ is the mass of the ground state and $s_0$ is the threshold
parameter. Above $s_0$, the spectral density at the hadron level is
replaced by the spectral density derived at the quark-gluon level.
Now the sum rule reads
\begin{equation}
f\mathrm{e}^{-\nicefrac{m^{2}}{M^{2}}}=\Pi^{\prime}\left(M^{2}\right)-\int_{s_{0}}^{\infty}\mathrm{e}^{-\nicefrac{s}{M^{2}}}\rho^{\text{OPE}}\left(s\right){\rm
ds}.\label{eq:sum rule}
\end{equation}
The usual numerical method in QCD sum rule analysis is to
differentiate Eq. (\ref{eq:sum rule}) with respect to
$\nicefrac{1}{M^{2}}$ and divide the resulting equation by Eq.
(\ref{eq:sum rule})
\begin{equation}
m^{2}=\frac{\int_{0}^{s_{0}}\mathrm{e}^{-\nicefrac{s}{M^{2}}}s\rho^{\text{OPE}}\left(s\right){\rm
ds}}{\int_{0}^{s_{0}}\mathrm{e}^{-\nicefrac{s}{M^{2}}}\rho^{\text{OPE}}\left(s\right){\rm
ds}}.
\end{equation}
One usually plots the variation of the mass versus $M^2$ and $s_0$
to find a working window.

However, the method described above can only be applied to the
ground states. In order to extract the resonance information of the
first radial excitation, we modify the above spectral density and
explicitly keep the pole of the first radial excitation in the
spectrum. Now the modified spectral density reads
\begin{equation}
\rho(s)\equiv\frac{1}{\pi}\mathrm{Im}\Pi\left(s\right)=f_{1}\delta\left(s-m^{2}\right)+f_{2}\delta\left(s-m^{\prime2}\right)+\rho_{continuum}\theta\left(s-s_{0}^{\prime}\right).\label{eq:
spectrum}
\end{equation}
To simply the numerical analysis, we use the zero width
approximation for both the ground state and first radial excitation.
The parameters $f_{1}$ and $f_{2}$ are related to the coupling
parameters while $m$ and $m'$ are the masses of the ground state and
the first radial excitation respectively. Now the sum rules read
\begin{eqnarray*}
 & \int\mathrm{e}^{-\nicefrac{s}{M^{2}}}\rho_{ground}\left(s\right)\mathrm{d}s+\int\mathrm{e}^{-\nicefrac{s}{M^{2}}}\rho_{\text{excitation}}\left(s\right)\mathrm{d}s+\int_{s_{0}}^{\infty}\mathrm{e}^{-\nicefrac{s}{M^{2}}}\rho_{\text{continuum}}\left(s\right)\mathrm{d}s\\
 & =\Pi^{\prime}\left(M^{2}\right)=\Pi^{\prime\text{perturbation}}\left(M^{2}\right)+\Pi^{\prime\text{condensates}}\left(M^{2}\right).
\end{eqnarray*}
The usual numerical method cannot be applied here because the
modified spectrum has two mass parameters. We use the least square
method \cite{Narison1984} to fit these masses and decay parameters. The detail of the
method are described below.

As usual in the sum rule analysis, one has to find an optimal
working interval of the Borel parameter $M^{2}$. The lower boundary
of $M^2$ is chosen to ensure the convergence of the operator product
expansion while the upper boundary is chosen to make the continuum
contribution remain subleading.

To get an optimal interval of the Borel parameter $M^{2}$, we set
\begin{equation}
\left|\frac{\int_{s_{0}}^{\infty}\mathrm{e}^{-\nicefrac{s}{M^{2}}}\rho_{\text{continuum}}\left(s\right)\mathrm{d}s}{\Pi^{\prime}\left(M^{2}\right)}\right|\leq\alpha_{1}
\end{equation}
which ensures that the continuum contribution remains subleading and
determines the upper boundary and
\begin{equation}
\left|\frac{\Pi^{\prime\text{condensates}}\left(M^{2}\right)}{\Pi^{\prime}\left(M^{2}\right)}\right|\leq\alpha_{2}
\end{equation}
which ensures that the OPE is reliable and determines the lower
boundary. The two boundaries determine the optimal interval of
$M^{2}$ for our numerical analysis.

The number $\alpha_{1}$ and $\alpha_{2}$ is chosen to ensure a
rational contribution of continuum and higher order OPE terms. For
the meson case, we set $\alpha_{1}=\alpha_{2}=\alpha$ to get a
reasonable interval of $M^{2}$. We use different values for
$\alpha_{1}$ and $\alpha_{2}$ in the nucleon case. Note that we
always try a smaller $\alpha$ in the excitation case since the
continuum contribution decreases as the threshold parameter $s_0$
increases. If no reasonable interval of $M^{2}$ can be got in any
way, the sum rule may not be appropriate in our numerical method.

We rewrite the sum rule as
\begin{eqnarray*}
 & \int\mathrm{e}^{-\nicefrac{s}{M^{2}}}\rho_{ground}\left(s\right)\mathrm{d}s+\int\mathrm{e}^{-\nicefrac{s}{M^{2}}}\rho_{excitation}\left(s\right)\mathrm{d}s\\
 & =g\left(M^{2},s_{0}\right)=\Pi^{\prime}\left(M^{2}\right)-\int_{s_{0}}^{\infty}\mathrm{e}^{-\nicefrac{s}{M^{2}}}\rho_{continuum}\left(s\right)\mathrm{d}s
\end{eqnarray*}
which separates the part of expression with physical parameters from
the part with just Borel parameter $M^{2}$ and threshold $s_{0}$.

With the above expression of $g\left(M^{2},s_{0}\right)$, we can
generate a series of points $\left\{\left(M_{i}^{2},
g\left(M_{i}^{2},s_{0}\right)\right)\right\} $ by choosing a set $\left\{
M_{i}^{2}\right\} $ within the optimal interval of $M^{2}$. We
uniformly choose $N$ points in the optimal interval of $M^{2}$. The
number $N$ is chosen to be $20$ or even larger.

With the sets $\left\{\left(M_{i}^{2},
g\left(M_{i}^{2},s_{0}\right)\right)\right\} $, we use the least-square
method which minimizes the sum of the squares of the difference
between the two sides of the sum rules
\begin{equation}
\sum_{i=1}^{N}\frac{\left|f_{1}\mathrm{e}^{-\frac{m^{2}}{M_{i}^{2}}}+f_{2}\mathrm{e}^{-\frac{m^{\prime2}}{M_{i}^{2}}}-g\left(M_{i}^{2},s_{0}\right)\right|^{2}}{N}=min\label{eq:least
square}
\end{equation}
to get the best fit of the resonance parameters of the ground state
and first radial excitation.

The masses of the ground states of the light mesons and nucleon are
measured precisely experimentally. The extracted masses from the
traditional QCD sum rule formalism with the one-pole approximation
agree with the experimental data very well. In our analysis we first
use the least square method to reproduce the resonance parameters of
the ground states. As expected, the resulting masses are consistent
with experimental data and those extracted from the traditional QCD
sum rule analysis.

Then we use the extracted masses of the ground states as inputs to
extract the resonance parameters of the radial excited states since
the less parameters in the fitting will cost less computing resource
and lead to relatively more stable results. Moreover, we do not fix
the masses of the ground states in Eq. (\ref{eq:least square}) in
our numerical analysis. Instead we allow them to vary around the
experimental central value within $\pm5\%$. In this way, we extract
the resonance parameters of the first radial excited states
numerically.

We analyze several light mesons and nucleon in the following
section. The sum rules of the light mesons can be found in the
pioneer paper \cite{Shifman1979}. The nucleon sum rule with the
radiative corrections can be found in Ref. \cite{Drukarev2009}. We
collect these sum rules in the appendix.

In our analysis we use the following values for the various
condensates and parameters
\cite{Olive2014,Ioffe2003,Shifman1979}: $\left\langle
\bar{q}q\right\rangle
\left(2\text{GeV}\right)=-\left(277_{-10}^{+12}\text{MeV}\right)^{3}$,
$\left\langle
0\right|m_{u}\bar{u}u+m_{d}\bar{d}d\left|0\right\rangle
=-\frac{1}{2}f_{\pi}^{2}m_{\pi}^{2}=-1.7\times10^{-4}{\rm GeV}^{4}
$, $m_{s}\left(2\text{GeV}\right)=(95\pm5)\text{MeV}$, $
\nicefrac{\left\langle \bar{s}s\right\rangle }{\left\langle
\bar{q}q\right\rangle }=0.8\pm0.3$, $ \left\langle
0\right|\frac{\alpha_{s}}{\pi}G_{\mu\nu}^{a}G_{\mu\nu}^{a}\left|0\right\rangle
=0.012_{-0.012}^{+0.006}{\rm GeV}^{4}$, $ \left\langle
0\right|\alpha_{s}\left(\bar{u}\gamma_{\alpha}\gamma_{5}t^{a}u-\bar{d}\gamma_{\alpha}\gamma_{5}t^{a}d\right)^{2}\left|0\right\rangle
=\frac{32}{9}\alpha_{s}\left\langle
0\right|\bar{q}q\left|0\right\rangle ^{2}\simeq6.5\times10^{-4}{\rm
GeV}^{4}$, $ \left\langle
0\right|\alpha_{s}\left(\bar{u}\gamma_{\alpha}\gamma_{5}t^{a}u-\bar{d}\gamma_{\alpha}\gamma_{5}t^{a}d\right)
\sum_{q=u,d,s}\bar{q}\gamma_{\alpha}t^{a}q\left|0\right\rangle
\simeq-\frac{32}{9}\alpha_{s}\left\langle
0\right|\bar{q}q\left|0\right\rangle ^{2}\simeq-6.5\times10^{-4}{\rm
GeV}^{4}$, $
\alpha_{s}\left(Q^{2}\right)=\nicefrac{4\pi}{\left(b\ln\left(\nicefrac{Q^{2}}{\Lambda^{2}}\right)\right)}
$, $\Lambda=0.1\text{GeV}$, $
\alpha_{s}\left(m_{Z}\right)=0.1184\pm0.0007 $,
$\alpha_{s}\left(1.5\text{GeV}\right)=0.353\pm0.006$.

\section{The $\rho$ meson}
\label{sec3rho}

The interpolating current for the $\rho$ meson is
\begin{equation}
j_{\mu}^{\left(\rho\right)}=\frac{1}{2}\left(\bar{u}\gamma_{\mu}u-\bar{d}\gamma_{\mu}d\right),
\end{equation}
and the resulting sum rule can be found in the appendix. The usual
single-pole spectral density reads
\begin{equation}
\rho^{\left(\rho\right)}\left(s\right)=6\pi^{2}f_{\rho}^{2}\delta\left(s-m_{\rho}^{2}\right)+\frac{3}{2}\left(1+\frac{\alpha_{s}\left(s\right)}{\pi}\right)\theta\left(s-s_{0}\right)
\end{equation}
We also need the double-pole spectral density
\begin{equation}
\rho^{\left(\rho\right)}\left(s\right)=6\pi^{2}f_{\rho}^{2}\delta\left(s-m_{\rho}^{2}\right)+6\pi^{2}f_{\rho^{\prime}}^{2}\delta\left(s-m_{\rho^{\prime}}^{2}\right)+\frac{3}{2}\left(1+\frac{\alpha_{s}\left(s\right)}{\pi}\right)\theta\left(s-s_{0}\right),
\end{equation}
where $f_{\rho}$ and $f_{\rho^{\prime}}$ are defined as
\begin{equation}
\left\langle 0\left|\bar{q}\gamma_{\mu}q\right|\rho\right\rangle
=m_{\rho}f_{\rho}\epsilon_{\mu},\left\langle
0\left|\bar{q}\gamma_{\mu}q\right|\rho^{\prime}\right\rangle
=m_{\rho^{\prime}}f_{\rho^{\prime}}\epsilon_{\mu}^{\prime}
\end{equation}
where $q=u,d$.

We first use the least square method and the traditional one-pole
spectrum representation with $\alpha=0.2$ and $N=40$ to extract the
mass and decay constant of the $\rho$ meson. The results are listed
in Table \ref{rho ground state}. The parameter $f_{1}$ is related to
the decay constant in Eq. (\ref{eq:usual spectrum}). The values of
"min" are the sum of the squares of the differences in Eq.
(\ref{eq:least square}). Only when the value of "min" is much
smaller than the parameters $f^2_{1}$, $f^2_{2}$ etc, the fit and
the extracted decay constants are reliable.

\begin{table}
\begin{center}
\begin{tabular}{c|ccccc}
\hline
$s_{0}${[}$\text{GeV}^{2}${]} & 1.2 & 1.3 & 1.4 & 1.5 & 1.6\tabularnewline
\hline
$M_{min}^{2}${[}$\text{GeV}^{2}${]} & 0.43 & 0.43 & 0.43 & 0.43 & 0.43\tabularnewline
$M_{max}^{2}${[}$\text{GeV}^{2}${]} & 0.74 & 0.82 & 0.88 & 0.94 & 1.00\tabularnewline
\hline
$m${[}GeV{]} & 0.74 & 0.75 & 0.75 & 0.76 & 0.77\tabularnewline
$f_{\rho}${[}MeV{]} & 187 & 190 & 193 & 197 & 201\tabularnewline
\hline
$f_{1}${[}$\text{GeV}^{2}${]} & 2.06 & 2.13 & 2.22 & 2.30 & 2.39\tabularnewline
min{[}$\text{GeV}^{4}${]} & $10^{-5}$ & $10^{-5}$ & $10^{-5}$ & $10^{-5}$ & $10^{-5}$\tabularnewline
\hline
\end{tabular}
\par\end{center}
\protect\caption{The mass and decay constant of the $\rho$ ground
state with $\alpha=0.2$ and $N=40$. \label{rho ground state}}
\end{table}

We collect the fitting results with the double-pole spectrum in
Table \ref{rho excitation}. Note that the parameter $m$ in Table
\ref{rho excitation} is the input to extract the information of the
excited state. We use $\alpha=0.1$ in this case. The threshold
$s_{0}$ plays the role of including the first radial excitation in
the spectrum while excluding the contribution from the higher
excitations. To check the consistency of our fitting and dependence
of our results on $s_0$, we vary $s_{0}$ in a range. A reliable
fitting requires that the mass $m^{\prime}$ and decay constant
$f_{\rho^{\prime}}$ of the first radial excitation should not vary
too much with $s_0$.

\begin{center}

\begin{table}
\begin{center}
\begin{tabular}{c|ccccc}
\hline
$s_{0}${[}$\text{GeV}^{2}${]} & 2.3 & 2.4 & 2.5 & 2.6 & 2.7\tabularnewline
\hline
$M_{min}^{2}${[}$\text{GeV}^{2}${]} & 0.50 & 0.50 & 0.50 & 0.50 & 0.50\tabularnewline
$M_{max}^{2}${[}$\text{GeV}^{2}${]} & 1.00 & 1.04 & 1.10 & 1.14 & 1.18\tabularnewline
\hline
$m${[}GeV{]}  & 0.76 & 0.76 & 0.76 & 0.76 & 0.76\tabularnewline
$m^{\prime}${[}GeV{]} & 1.24 & 1.29 & 1.35 & 1.38 & 1.40\tabularnewline
$f_{\rho}${[}MeV{]}  & 196 & 197 & 198 & 198 & 198\tabularnewline
$f_{\rho^{\prime}}${[}MeV{]} & 130 & 141 & 152 & 161 & 170\tabularnewline
\hline
$f_{1}${[}$\text{GeV}^{2}${]} & 2.3 & 2.3 & 2.3 & 2.3 & 2.3\tabularnewline
$f_{2}[\mbox{GeV}^{2}]$ & 1.0 & 1.2 & 1.4 & 1.5 & 1.7\tabularnewline
min{[}$\text{GeV}^{4}${]} & $10^{-6}$ & $10^{-6}$ & $10^{-6}$ & $10^{-6}$ & $10^{-6}$\tabularnewline
\hline
\end{tabular}
\par\end{center}

\protect\caption{Masses and decay constants of the $\rho$ ground
state and first radial excitation with $\alpha=0.1$ and $N=40$.
\label{rho excitation}}
\end{table}
\par\end{center}

From Table \ref{rho ground state} we have
\begin{equation}
m=\left(0.76\pm0.01\right)\text{GeV},f_{\rho}=\left(194\pm6\right)\text{MeV}
\end{equation}
which agrees with the $\rho$ meson mass from PDG $m=0.77$GeV
\cite{Olive2014} and the experimental measurement of the $\rho$ meson
decay constant \cite{Becirevic2003}
\begin{equation}
f_{\rho}^{\text{exp}}\simeq216\left(5\right)\text{MeV}.
\end{equation}
In order to reduce the dependence on the threshold parameter $s_0$,
the extracted values of $m$ and $f_{\rho}$ are the average values of
the numerical values in Table \ref{rho ground state}. From Table
\ref{rho excitation} we have
\begin{equation}
m^{\prime}=\left(1.33\pm0.07\right)\text{GeV},f_{\rho}=\left(197\pm1\right)\text{MeV},f_{\rho^{\prime}}=\left(151\pm16\right)\text{MeV}.
\end{equation}
From PDG, the mass of the first radial excitation is
$m^{\prime}=1.47$ GeV and its width is $\Gamma=0.40$ GeV. Our
extracted $\rho^{\prime}$ mass is consistent with the experimental
data. At present, the decay constant of $\rho^{\prime}$ has not been
measured yet.

\section{The $\pi$ and $A_1$ mesons}
\label{sec3pi}

We adopt the axial current for the pion and $A_1$ mesons
\begin{equation}
j_{5\mu}^{A_1}=\bar{u}\gamma_{\mu}\gamma_{5}d,
\end{equation}
and the resulting sum rule can be found in the appendix. Besides the
$a_1$ pole, the pion also contributes to this sum rule due to the
partial conservation of the axial vector current. As a Goldstone
boson, the pion mass is tiny. Especially in the sum rule analysis,
$m_\pi^2$ is much much less than the Borel parameter $M^2$. We can
safely ignore the pion mass and let it be zero in the numerical
analysis.

The usual spectrum representation is
\begin{equation}
\rho\left(s\right)=\pi f_{\pi}^{2}\delta\left(s\right)+\pi
f_{A_{1}}^{2}\delta\left(s-m_{A_{1}}^{2}\right)+\frac{1}{4\pi}\left(1+\frac{\alpha_{s}\left(s\right)}{\pi}\right)\theta\left(s-s_{0}\right).
\end{equation}
Our modified spectrum representation reads
\begin{equation}
\rho^{\left(\pi\right)}\left(s\right)=\pi
f_{\pi}^{2}\delta\left(s\right)+\pi
f_{\pi^{\prime}}^{2}\delta\left(s-m_{\pi^{\prime}}^{2}\right)+\pi
f_{A_{1}}^{2}\delta\left(s-m_{A_{1}}^{2}\right)+\frac{1}{4\pi}\left(1+\frac{\alpha_{s}\left(s\right)}{\pi}\right)\theta\left(s-s_{0}\right),
\end{equation}
where $f_{\pi}$, $f_{\pi^{\prime}}$ $f_{A_{1}}$ are defined as
\begin{equation}
\left\langle 0\left|j_{\mu}^{\pi}\right|\pi\right\rangle
=if_{\pi}p_{\mu},\left\langle
0\left|j_{\mu}^{\pi}\right|\pi^{\prime}\right\rangle
=if_{\pi^{\prime}}p_{\mu}^{\prime},\left\langle
0\left|j_{\mu}^{\pi}\right|A_{1}\right\rangle
=m_{A_{1}}f_{A_{1}}\epsilon_{\mu}^{\prime}.
\end{equation}

In the fitting, we use the least square method and the traditional
spectrum representation with $\alpha=0.3$ and $N=80$ to extract the
$A_1$ mass and decay constant. The results are listed in Table
\ref{pion ground state}.

\begin{center}
\begin{table}
\begin{center}
\begin{tabular}{c|ccccc}
\hline
$s_{0}${[}$\text{GeV}^{2}${]} & 1.3 & 1.40 & 1.50 & 1.60 & 1.70\tabularnewline
\hline
$M_{min}^{2}${[}$\text{GeV}^{2}${]} & 0.52 & 0.52 & 0.52 & 0.52 & 0.52\tabularnewline
$M_{max}^{2}${[}$\text{GeV}^{2}${]} & 1.16 & 1.20 & 1.28 & 1.36 & 1.44\tabularnewline
\hline
$m_{A_{1}}${[}GeV{]} & 1.14 & 1.18 & 1.22 & 1.26 & 1.28\tabularnewline
$f_{\pi}${[}MeV{]} & 134 & 135 & 136 & 137 & 137\tabularnewline
$f_{A_{1}}${[}MeV{]} & 124 & 139 & 153 & 166 & 175\tabularnewline
\hline
$f_{1}${[}$\text{GeV}^{2}${]} & 0057 & 0.057 & 0.058 & 0.059 & 0.059\tabularnewline
$f_{2}[\mbox{GeV}^{2}]$ & 0.048 & 0.060 & 0.074 & 0.087 & 0.096\tabularnewline
min{[}$\text{GeV}^{4}${]} & $10^{-7}$ & $10^{-7}$ & $10^{-7}$ & $10^{-8}$ & $10^{-8}$\tabularnewline
\hline
\end{tabular}
\par\end{center}
\protect\caption{The mass and decay constant of the $A_1$ meson. We
use the least square method and the traditional spectrum
representation with $\alpha=0.3$ and $N=80$. \label{pion ground
state}}
\end{table}
\par\end{center}

In order to extract the resonance parameters of the first excitation
of the pion meson, we employ the modified spectrum and allow
$f_{A_{1}}$ and $m_{A_{1}}$ to vary around the experimental data
within $\pm5\%$. The numerical results are listed in Table \ref{pion
excitation}.

\begin{center}

\begin{table}
\begin{center}
\begin{tabular}{c|ccccccc}
\hline
$s_{0}${[}$\text{GeV}^{2}${]} & 2.0 & 2.1 & 2.2 & 2.3 & 2.4 & 2.5 & 2.6\tabularnewline
\hline
$M_{min}^{2}${[}$\text{GeV}^{2}${]} & 0.63 & 0.63 & 0.63 & 0.63 & 0.63 & 0.63 & 0.63\tabularnewline
$M_{max}^{2}${[}$\text{GeV}^{2}${]} & 1.28 & 1.34 & 1.38 & 1.44 & 1.50 & 1.56 & 1.62\tabularnewline
\hline
$m_{A_{1}}${[}GeV{]} & 1.29 & 1.29 & 1.29 & 1.29 & 1.29 & 1.29 & 1.29\tabularnewline
$m_{\pi}^{\prime}${[}GeV{]} & 1.34 & 1.36 & 1.31 & 1.34 & 1.41 & 1.43 & 1.46\tabularnewline
$f_{\pi}${[}MeV{]} & 121 & 122 & 123 & 123 & 124 & 125 & 126\tabularnewline
$f_{A_{1}}${[}MeV{]} & 248 & 248 & 248 & 248 & 248 & 248 & 248\tabularnewline
$f_{\pi^{\prime}}${[}MeV{]} & 0.2 & 0.3 & 0.1 & 2.3 & 0.2 & 0.7 & 0.1\tabularnewline
\hline
$f_{1}${[}$\text{GeV}^{2}${]} & 0.05 & 0.05 & 0.05 & 0.05 & 0.05 & 0.05 & 0.05\tabularnewline
$f_{2}[\mbox{GeV}^{2}]$ & 0.19 & 0.19 & 0.19 & 0.19 & 0.19 & 0.19 & 0.19\tabularnewline
$f_{3}[\mbox{GeV}^{2}]$ & $10^{-7}$ & $10^{-7}$ & $10^{-8}$ & $10^{-5}$ & $10^{-7}$ & $10^{-6}$ & $10^{-8}$\tabularnewline
min{[}$\text{GeV}^{4}${]} & $10^{-5}$ & $10^{-5}$ & $10^{-5}$ & $10^{-5}$ & $10^{-5}$ & $10^{-5}$ & $10^{-5}$\tabularnewline
\hline
\end{tabular}
\par\end{center}
\protect\caption{Masses and decay constants of the $A_1$ ground
state and the first radial excitation of the pion with $\alpha=0.2$
and $N=80$. \label{pion excitation}}
\end{table}

\par\end{center}

From Table \ref{pion ground state} we have
\begin{equation}
m_{A_{1}}=\left(1.22\pm0.06\right)\text{GeV},f_{\pi}=\left(135\pm1\right)\text{MeV},f_{A_{1}}=\left(151\pm20\right)\text{MeV}.
\end{equation}
From PDG we have $m_{A_{1}}=1.23$ GeV and $\Gamma_{A_{1}}=0.40$ GeV.
We note that the $A_1$ mass from the fitting is in
rough agreement with the experimental data. The extracted pion decay
constant agrees with the experimental data \cite{Olive2014}
\begin{equation}
f_{\pi}^{\text{exp}}=130\text{MeV}.
\end{equation}
However, the extracted $A_1$ decay constant is only half of the
experimental data \cite{Wingate1995}
\begin{equation}
f_{A_{1}}^{\text{exp}}=254\left(20\right)\text{MeV}.
\end{equation}

To extract the first radial excitation of the pion meson, we use the
experimental data of the $A_1$ decay constant as input in the
numerical analysis. The results are collected in Table \ref{pion
excitation}. We have
\begin{equation}
m_{\pi^{\prime}}=\left(1.38\pm0.06\right)\text{GeV},f_{\pi}=\left(123\pm1\right)\text{MeV},f_{\pi^{\prime}}=\left(0.6\pm0.8\right)\text{MeV}.
\end{equation}
The resulting mass of the pion radial excitation agrees with the PDG
value very nicely: $m_{\pi^{\prime}}=1.30$ GeV and
$\Gamma_{\pi^{\prime}}=0.40$ GeV \cite{Olive2014}. Note that the extracted
numerical value of $f_{\pi^{\prime}}$ is not reliable since the
parameter $f^2_{3}$ is even smaller than the "min". In this case, we
may get a upper bound
\begin{equation}
\left|f_{3}\right|<\sqrt{min}\sim0.0032\mbox{GeV}^{2}
\end{equation}
Accordingly, we get the upper bound for $f_{\pi^{\prime}}$
\begin{equation}
f_{\pi^{\prime}}<0.032\mbox{GeV}
\end{equation}
If the value of $f_{\pi^{\prime}} $ is larger than 0.032 GeV, we
should be able to extract its value through the least square fitting
method.

In other words, our numerical analysis demonstrates that the decay
constant of the pion radial excitation $\pi^{\prime}$ is much
smaller than the pion decay constant around 130 MeV. This
interesting fact was also noticed by previous theoretical work
including lattice simulations
\cite{Andrianov1998,Elias1997,Maltman2002,McNeile2006,Volkov1997,Holl2004,Holl2005,Qin2012,Narison2014,Kataev1983,Gorishnii1984}.
In fact, the suppression of the $\pi^{\prime}$ decay constant is a
consequence of the chiral symmetry breaking. In the chiral limit,
the decay constants of the pion and its radial excitations satisfy
the following relation \cite{Dominguez1977}
\begin{equation}
f_{\pi_{n}}m_{\pi_{n}}^{2}=0,
\end{equation}
where $m_{\pi_{n}}$ ($n\geq 1$) is the mass of the pion radial
excitation. The pion ground state is massless in the chiral limit as
a Goldstone boson, hence its decay constant can be large and
nonzero. For the pion radial excitation, its mass is large and
nonzero. Therefore its decay constant has to vanish, i.e.,
$f_{\pi_{1}}=0$.

\section{The $K^{*}$ meson}
\label{sec3kaon}

The interpolating current for the $K^{*}$ meson is
\begin{equation}
j_{\mu}^{\left(K^{*}\right)}=\bar{u}\gamma_{\mu}s
\end{equation}
and the resulting sum rule can be found in the appendix. The usual
single-pole spectral density reads
\begin{equation}
\mathrm{\rho}\left(s\right)=\pi
f_{K^{*}}^{2}\delta\left(s-m_{K^{*}}^{2}\right)+\frac{1}{4\pi}\left(1+\frac{\alpha_{s}\left(s\right)}{\pi}\right)\theta\left(s-s_{0}\right).
\end{equation}
Our modified spectrum representation reads
\begin{equation}
\rho^{\left(K^{*}\right)}\left(s\right)=\pi
f_{K^{*}}^{2}\delta\left(s-m_{K^{*}}^{2}\right)+\pi
f_{K^{*\prime}}^{2}\delta\left(s-m_{K^{*\prime}}^{2}\right)+\frac{1}{4\pi}\left(1+\frac{\alpha_{s}\left(s\right)}{\pi}\right)\theta\left(s-s_{0}\right),
\end{equation}
where $f_{K^{*}}$ and $f_{K^{*\prime}}$ are defined as
\begin{equation}
\left\langle
0\left|j_{\mu}^{\left(K^{*}\right)}\right|K^{*}\right\rangle
=m_{K^{*}}f_{K^{*}}\epsilon_{\mu},\left\langle
0\left|j_{\mu}^{\left(K^{*}\right)}\right|K^{*\prime}\right\rangle
=m_{K^{*\prime}}f_{K^{*\prime}}\epsilon_{\mu}^{\prime}.
\end{equation}

\begin{center}
\begin{table}
\begin{center}
\begin{tabular}{c|ccccc}
\hline
$s_{0}${[}$\text{GeV}^{2}${]} & 1.4 & 1.5 & 1.6 & 1.6 & 1.8\tabularnewline
\hline
$M_{min}^{2}${[}$\text{GeV}^{2}${]} & 0.63 & 0.63 & 0.63 & 0.63 & 0.63\tabularnewline
$M_{max}^{2}${[}$\text{GeV}^{2}${]} & 1.10 & 1.18 & 1.28 & 1.36 & 1.44\tabularnewline
\hline
$m${[}GeV{]} & 0.88 & 0.89 & 0.90 & 0.90 & 0.91\tabularnewline
$f_{K^{*}}${[}MeV{]} & 202 & 206 & 210 & 215 & 219\tabularnewline
\hline
$f_{1}${[}$\text{GeV}^{2}${]} & 0.13 & 0.13 & 0.14 & 0.14 & 0.15\tabularnewline
min{[}$\text{GeV}^{4}${]} & $10^{-7}$ & $10^{-7}$ & $10^{-7}$ & $10^{-7}$ & $10^{-7}$\tabularnewline
\hline
\end{tabular}
\par\end{center}
\protect\caption{The mass and decay constant of the $K^{*}$ ground
state. We use the least square method and the traditional spectrum
representation with $\alpha=0.3$ and $N=20$. \label{K star ground
state}}
\end{table}

\par\end{center}

\begin{table}
\begin{center}
\begin{tabular}{c|cccccc}
\hline
$s_{0}${[}$\text{GeV}^{2}${]} & 2.3 & 2.4 & 2.5 & 2.6 & 2.7 & 2.8\tabularnewline
\hline
$M_{min}^{2}${[}$\text{GeV}^{2}${]} & 0.63 & 0.63 & 0.63 & 0.63 & 0.63 & 0.63\tabularnewline
$M_{max}^{2}${[}$\text{GeV}^{2}${]} & 1.40 & 1.46 & 1.52 & 1.58 & 1.64 & 1.70\tabularnewline
\hline
$m${[}GeV{]} & 0.89 & 0.89 & 0.89 & 0.89 & 0.89 & 0.89\tabularnewline
$m^{\prime}${[}GeV{]} & 1.22 & 1.25 & 1.27 & 1.33 & 1.29 & 1.37\tabularnewline
$f_{K^{*}}${[}MeV{]} & 200 & 201 & 202 & 200 & 207 & 207\tabularnewline
$f_{K^{*\prime}}${[}MeV{]} & 139 & 146 & 153 & 162 & 159 & 172\tabularnewline
\hline
$f_{1}${[}$\text{GeV}^{2}${]} & 0.13 & 0.13 & 0.13 & 0.12 & 013 & 0.13\tabularnewline
$f_{2}[\mbox{GeV}^{2}]$ & 0.06 & 0.07 & 0.08 & 0.08 & 0.08 & 0.09\tabularnewline
min{[}$\text{GeV}^{4}${]} & $10^{-8}$ & $10^{-8}$ & $10^{-8}$ & $10^{-8}$ & $10^{-8}$ & $10^{-8}$\tabularnewline
\hline
\end{tabular}
\par\end{center}
\protect\caption{Masses and decay constants of the $K^{*}$ ground
state and the first radial excitation with $\alpha=0.2$ and $N=20$. \label{K star excitation}}
\end{table}

The results from the first spectrum representation are listed in
Table \ref{K star ground state} and those from the modified spectrum
are listed in Table \ref{K star excitation}. From Table \ref{K star
ground state} we have
\begin{equation}
m=\left(0.89\pm0.01\right)\text{GeV},f_{K^{*}}=\left(210\pm7\right)\text{MeV}.
\end{equation}
From Table \ref{K star excitation} we have
\begin{equation}
m^{\prime}=\left(1.28\pm0.06\right)\text{GeV},f_{K^{*}}=\left(203\pm3\right)\text{MeV},f_{K^{*\prime}}=\left(155\pm11\right)\text{MeV},
\end{equation}
where m is an input parameter in Table \ref{K star excitation}. The
decay constant of the $K^{*}$ was measured to be
\cite{Becirevic2003}
\begin{equation}
f_{K^{*}}^{\text{exp}}\simeq217\text{MeV}.
\end{equation}
From PDG, the mass and width of the $K^{*\prime}$ are
$m^{\prime}=1.41$ GeV, $\Gamma=0.232$ GeV respectively. Clearly our
extracted $f_{K^{*}}$ from both fitting agrees with the data. The
extracted $m^{\prime}$ is also consistent with data.

\section{The $\varphi$ meson}
\label{sec3phi}

The interpolating current for the $\varphi$ meson is
\begin{equation}
j_{\mu}^{\left(\varphi\right)}=-\frac{1}{3}\bar{s}\gamma_{\mu}s,
\end{equation}
and the resulting sum rule can be found in the appendix. The usual
spectrum representation is
\begin{equation}
\rho\left(s\right)=\frac{1}{9}\pi
f_{\varphi}^{2}\delta\left(s-m_{\varphi}^{2}\right)+\frac{1}{36\pi}\left(1+\frac{\alpha_{s}\left(s\right)}{\pi}\right)\theta\left(s-s_{0}\right).
\end{equation}
We also use the modified spectrum representation
\begin{equation}
\rho^{\left(\varphi\right)}\left(s\right)=\frac{1}{9}\pi
f_{\varphi}^{2}\delta\left(s-m_{\varphi}^{2}\right)+\frac{1}{9}\pi
f_{\varphi^{\prime}}^{2}\delta\left(s-m_{\varphi^{\prime}}^{2}\right)+\frac{1}{36\pi}\left(1+\frac{\alpha_{s}\left(s\right)}{\pi}\right)\theta\left(s-s_{0}\right)
\end{equation}
where $f_{\varphi}$ and $f_{\varphi^{\prime}}$ are defined as
\begin{equation}
\left\langle 0\left|\bar{s}\gamma^{\mu}s\right|\varphi\right\rangle
=m_{\varphi}f_{\varphi}\epsilon_{\mu},\left\langle
0\left|\bar{s}\gamma^{\mu}s\right|\varphi^{\prime}\right\rangle
=m_{\varphi^{\prime}}f_{\varphi^{\prime}}\epsilon_{\mu}^{\prime}.
\end{equation}

We use the least square method and the traditional spectrum
representation with $N=20$. Note that there does not exist a working
interval of $M^{2}$ for $\alpha=0.2$. So we use $\alpha=0.3$ here.
The results from the first spectrum representation are listed in
Table \ref{phi ground state} and those from the modified spectrum
are listed in Table \ref{phi excitation}, where m is the input
parameter in Table \ref{phi excitation}.

\begin{center}
\begin{table}
\begin{center}
\begin{tabular}{c|cccccc}
\hline
$s_{0}${[}$\text{GeV}^{2}${]} & 1.7 & 1.8 & 1.9 & 2.0 & 2.1 & 2.2\tabularnewline
\hline
$M_{min}^{2}${[}$\text{GeV}^{2}${]} & 0.87 & 0.87 & 0.87 & 0.87 & 0.87 & 0.87\tabularnewline
$M_{max}^{2}${[}$\text{GeV}^{2}${]} & 1.66 & 1.78 & 1.90 & 2.00 & 2.12 & 2.24\tabularnewline
\hline
$m${[}GeV{]} & 1.02 & 1.03 & 1.03 & 1.04 & 1.05 & 1.06\tabularnewline
$f_{\varphi}${[}Mev{]} & 217 & 221 & 226 & 231 & 236 & 240\tabularnewline
\hline
$f_{1}${[}$\text{GeV}^{2}${]} & 0.016 & 0.017 & 0.018 & 0.019 & 0.019 & 0.020\tabularnewline
min{[}$\text{GeV}^{4}${]} & $10^{-9}$ & $10^{-9}$ & $10^{-9}$ & $10^{-9}$ & $10^{-9}$ & $10^{-9}$\tabularnewline
\hline
\end{tabular}
\par\end{center}
\protect\caption{The mass and decay constant of the $\varphi$ ground
state. We use the least square method and the traditional spectrum
representation with $N=20$ and $\alpha=0.3$ here. \label{phi ground
state}}
\end{table}

\par\end{center}

\begin{center}
\begin{table}
\begin{center}
\begin{tabular}{c|ccccccc}
\hline
$s_{0}${[}$\text{GeV}^{2}${]} & 3.4 & 3.5 & 3.6 & 3.7 & 3.8 & 3.9 & 4.0\tabularnewline
\hline
$M_{min}^{2}${[}$\text{GeV}^{2}${]} & 1.15 & 1.15 & 1.15 & 1.15 & 1.15 & 1.15 & 1.15\tabularnewline
$M_{max}^{2}${[}$\text{GeV}^{2}${]} & 2.02 & 2.08 & 2.16 & 2.22 & 2.28 & 2.36 & 2.42\tabularnewline
\hline
$m${[}GeV{]} & 1.00 & 1.00 & 1.00 & 1.00 & 1.00 & 1.00 & 1.00\tabularnewline
$m^{\prime}${[}GeV{]} & 1.45 & 1.64 & 1.52 & 1.55 & 1.62 & 1.50 & 1.51\tabularnewline
$f_{\varphi}${[}MeV{]}  & 203 & 221 & 210 & 213 & 218 & 203 & 202\tabularnewline
$f_{\varphi^{\prime}}${[}MeV{]} & 215 & 215 & 222 & 226 & 230 & 240 & 246\tabularnewline
\hline
$f_{1}${[}$\text{GeV}^{2}${]} & 0.014 & 0.017 & 0.015 & 0.016 & 0.017 & 0.014 & 0.014\tabularnewline
$f_{2}[\mbox{GeV}^{2}]$ & 0.016 & 0.016 & 0.017 & 0.018 & 0.018 & 0.020 & 0.021\tabularnewline
min{[}$\text{GeV}^{4}${]} & $10^{-10}$ & $10^{-10}$ & $10^{-10}$ & $10^{-10}$ & $10^{-10}$ & $10^{-10}$ & $10^{-9}$\tabularnewline
\hline
\end{tabular}
\par\end{center}
\protect\caption{Masses and decay constants of the $\varphi$ ground
state and the first radial excitation with $\alpha=0.2$ and $N=20$.
\label{phi excitation}}
\end{table}
\par\end{center}

From PDG, the mass and width of the $\varphi$ ground state are
$m=1.020$ GeV and $\Gamma=0.004$ GeV while $m^{\prime}=1.68$ GeV,
$\Gamma=0.20$ GeV for first radial excitation. The decay constant of
ground state was measured to be \cite{Becirevic2003}
\begin{equation}
f_{\varphi}^{\text{exp}}=233\text{MeV}.
\end{equation}
From Table \ref{phi ground state} we have
\begin{equation}
m=\left(1.04\pm0.02\right)\text{GeV},f_{\varphi}=\left(229\pm9\right)\text{MeV}.
\end{equation}
From Table \ref{phi excitation} we have
\begin{equation}
m^{\prime}=\left(1.54\pm0.07\right)\text{GeV},f_{\varphi}=\left(210\pm8\right)\text{MeV},f_{\varphi^{\prime}}=\left(228\pm11\right)\text{MeV}.
\end{equation}
The decay constant of the $\varphi$ meson from both fittings agrees
with the data very well while the extracted mass of the first radial
excitation is in rough agreement with the data.

\section{The nucleon}
\label{sec3nucleon}

The interpolating current for the nucleon is
\begin{equation}
\eta=\epsilon^{abc}\left[u^{aT}Cd^{b}\right]\gamma^{5}u^{c}-\epsilon^{abc}\left[u^{aT}C\gamma^{5}d^{b}\right]u^{c}
\end{equation}
and the resulting sum rule \cite{Drukarev2009} can be found in the
appendix. The usual spectrum representation for the nucleon is
\begin{equation}
\rho^{\left(N\right)}\left(s\right)=\beta_{N}^{2}\delta\left(s-m^{2}\right)+\rho_{continuum}\left(s\right)\theta\left(s-s_{0}\right)
\end{equation}
where
\begin{eqnarray}
\rho_{\text{continuum}}\left(s\right) & = & \frac{1}{\pi}\mathrm{Im}\Pi\left(s\right)\nonumber \\
 & = & \frac{s^{2}}{4\left(2\pi\right)^{4}}\left(1+\frac{71}{12}\frac{\alpha_{s}}{\pi}-\frac{\alpha_{s}}{\pi}\mathrm{ln}\frac{s}{\mu^{2}}\right)+\frac{1}{\left(2\pi\right)^{2}}\frac{1}{8}\left\langle \frac{\alpha_{s}}{\pi}G^{2}\right\rangle -\frac{2\left\langle \bar{q}q\right\rangle ^{2}}{9}\frac{\alpha_{s}}{\pi}\frac{1}{s}
\end{eqnarray}

We also use the modified spectrum representation
\begin{equation}
\rho\left(s\right)=\beta_{N}^{2}\delta\left(s-m^{2}\right)+\beta_{N^{\prime}}^{2}\delta\left(s-m^{\prime2}\right)+\rho_{continuum}\left(s\right)\theta\left(s-s_{0}\right),
\end{equation}
where $\beta_{N}^{2}=32\pi^{4}\lambda_{N}^{2}$,
$\beta_{N^{\prime}}^{2}=32\pi^{4}\lambda_{N^{\prime}}^{2}$ and
$\lambda_{N}$ is the overlapping amplitude of the interpolating
current with the nucleon state.

\begin{center}
\begin{table}
\begin{center}
\begin{tabular}{c|ccccc}
\hline
$s_{0}${[}$\text{GeV}^{2}${]} & 1.80 & 1.85 & 1.90 & 1.95 & 2.0\tabularnewline
\hline
$M_{min}^{2}${[}$\text{GeV}^{2}${]} & 0.7 & 0.7 & 0.7 & 0.7 & 0.7\tabularnewline
$M_{max}^{2}${[}$\text{GeV}^{2}${]} & 1.52 & 1.54 & 1.58 & 1.62 & 1.64\tabularnewline
\hline
$m${[}GeV{]} & 0.89 & 0.91 & 0.93 & 0.94 & 0.96\tabularnewline
\hline
$\beta_{N}^{2}$ & 1.9 & 2.0 & 2.1 & 2.2 & 2.4\tabularnewline
min & $10^{-4}$ & $10^{-4}$ & $10^{-4}$ & $10^{-4}$ & $10^{-4}$\tabularnewline
\hline
\end{tabular}
\par\end{center}
\protect\caption{The mass of the nucleon ground state with
$\alpha_{1}=0.8$, $\alpha_{2}=0.4$ and $N=20$. \label{nucleon ground
state}}
\end{table}

\par\end{center}

\begin{table}
\begin{center}
\begin{tabular}{c|ccccccc}
\hline
$s_{0}${[}$\text{GeV}^{2}${]} & 2.1 & 2.15 & 2.20 & 2.25 & 2.30 & 2.35 & 2.40\tabularnewline
\hline
$M_{min}^{2}${[}$\text{GeV}^{2}${]} & 0.83 & 0.83 & 0.83 & 0.83 & 0.83 & 0.83 & 0.83\tabularnewline
$M_{max}^{2}${[}$\text{GeV}^{2}${]} & 1.36 & 1.38 & 1.40 & 1.42 & 1.44 & 1.48 & 1.50\tabularnewline
\hline
$m${[}GeV{]} & 0.929 & 0.929 & 0.929 & 0.929 & 0.929 & 0.929 & 0.929\tabularnewline
$m^{\prime}${[}GeV{]} & 1.45 & 1.47 & 1.48 & 1.50 & 1.52 & 1.53 & 1.55\tabularnewline
\hline
$\beta_{N}^{2}$ input & 2.1 & 2.1 & 2.1 & 2.2 & 2.2 & 2.2 & 2.2\tabularnewline
$\beta_{N^{\prime}}^{2}$ & 0.67 & 0.86 & 1.06 & 1.28 & 1.50 & 1.76 & 2.00\tabularnewline
min & $10^{-5}$ & $10^{-5}$ & $10^{-5}$ & $10^{-5}$ & $10^{-5}$ & $10^{-5}$ & $10^{-5}$\tabularnewline
\hline
\end{tabular}
\par\end{center}
\protect\caption{The masses of the nucleon ground state and the
first radial excitation with $\alpha_{1}=0.7$, $\alpha_{2}=0.3$ and
$N=20$.\label{nucleon excitation}}
\end{table}

The results from the first spectrum representation are listed in
Table \ref{nucleon ground state} and those from the modified
spectrum are listed in Table \ref{nucleon excitation}. To get stable
results, we have used the nucleon mass and $\beta_{N}^{2}=2.1$ from
Table \ref{nucleon ground state} as input in the numerical analysis
of the first radial excitation. From PDG, the nucleon mass is
$m=0.938$ GeV while the mass and width of its first radial
excitation are $m^{\prime}=1.44$ GeV and $\Gamma^{\prime}=0.300$
GeV. From Table \ref{nucleon ground state} we have
\begin{equation}
m=\left(0.93\pm0.03\right)\text{GeV}.
\end{equation}
From Table \ref{nucleon excitation} we have
\begin{equation}
m^{\prime}=\left(1.50\pm0.04\right)\text{GeV}
\end{equation}
which is in rough agreement with the data.

\section{Summary}
\label{sec4}

In short summary, we have attempted to extract the masses of the
first radial excited states of the light mesons and nucleon. In our
modified hadronic spectral density, we explicitly keep the pole of
the first radial excited states together with the ground state.
Requiring that the operator product expansion converge and the
continuum contribution be subleading leads to the optimal working
interval of the Borel parameter $M^{2}$. Then a series of "data"
points (or pseudo-data points) were produced within this working
interval of $M^{2}$. Using the usual one-pole spectral density, we
can extract the mass of the ground state with the least square
fitting method, which agrees with the experimental data. Then we use
these "data" points and the mass of the ground state as input
parameters to extract the mass and the decay constant of the first
radial excited state by the least square method, which are in good
agreement with the available data.

The QCD sum rule method has its inherent accuracy limit due to the
various approximations adopted within this framework, such as the
truncation of the the OPE series of the correlation function, the
assumption of the quark-hadron duality, the omission of the decay
width in the spectral density, the factorization of the four quark
condensates and the uncertainties of the values of the various
condensates etc. In our analysis we only include the uncertainty
from the fitting using the least square method itself. The least
square method with the modified spectrum representation allows us to
extract useful information of the first radial excitations, which
depends on the accuracy of the sum rules. It will be very
interesting to explore whether such a formalism can be applied to
the other hadrons.

\section*{ACKNOWLEDGMENTS}

This project is supported by the National Natural Science Foundation
of China under Grant No. 11261130311.

\clearpage
\appendix

\section*{QCD sum rules of the light mesons and nucleon}

For the $\rho$ meson
\begin{eqnarray}
 & \int\mathrm{d}s\,\mathrm{e}^{-\nicefrac{s}{M^{2}}}\rho\left(s\right)=\frac{3}{2}M^{2}\biggl[1+\frac{\alpha_{s}\left(M\right)}{\pi}+\frac{4\pi^{2}\left\langle 0\left|m_{u}\bar{u}u+m_{d}\bar{d}d\right|0\right\rangle }{M^{4}}\nonumber \\
 & +\frac{1}{3}\pi^{2}\frac{\left\langle 0\left|\frac{\alpha_{s}}{\pi}G_{\mu\nu}^{a}G_{\mu\nu}^{a}\right|0\right\rangle }{M^{4}}-2\pi^{3}\frac{\left\langle 0\left|\alpha_{s}\left(\bar{u}\gamma_{\alpha}\gamma_{5}t^{a}u-\bar{d}\gamma_{\alpha}\gamma_{5}t^{a}d\right)^{2}\right|0\right\rangle }{M^{6}}\\
 & -\frac{4}{9}\pi^{3}\frac{\left\langle 0\left|\alpha_{s}\left(\bar{u}\gamma_{\alpha}t^{a}u+\bar{d}\gamma_{\alpha}t^{a}d\underset{q=u,d,s}{\sum}\bar{q}\gamma_{\alpha}t^{a}q\right)\right|0\right\rangle }{M^{6}}\biggr]\nonumber
\end{eqnarray}
For the $\pi$ meson
\begin{eqnarray}
 & \int\mathrm{e}^{-\nicefrac{s}{M^{2}}}\rho\left(s\right)\mathrm{d}s=\frac{M^{2}}{4\pi}\biggl[1+\frac{\alpha_{s}\left(M\right)}{\pi}+\frac{1}{3}\pi^{2}\frac{\left\langle 0\left|\frac{\alpha_{s}}{\pi}G_{\mu\nu}^{a}G_{\mu\nu}^{a}\right|0\right\rangle }{M^{4}}\nonumber \\
 & +\frac{4\pi^{3}\alpha_{s}\left\langle 0\left|\bar{u}\gamma_{\alpha}\gamma_{5}t^{a}d\bar{d}\gamma_{\alpha}\gamma_{5}t^{a}u\right|0\right\rangle }{M^{6}}\\
 & -\frac{4}{9}\pi^{3}\alpha_{s}\frac{\left\langle 0\left|\left(\bar{u}\gamma_{\alpha}t^{a}u+\bar{d}\gamma_{\alpha}t^{a}d\underset{q=u,d,s}{\sum}\bar{q}\gamma_{\alpha}t^{a}q\right)\right|0\right\rangle }{M^{6}}\biggr]\nonumber
\end{eqnarray}
For the $K^{*}$ meson
\begin{eqnarray}
 & \int\mathrm{d}s\,\mathrm{e}^{-\nicefrac{s}{M^{2}}}\rho^{\left(K^{*}\right)}\left(s\right)=\frac{M^{2}}{4\pi}\biggl[1+\frac{\alpha_{s}\left(M\right)}{\pi}+\frac{14}{3}\frac{\pi^{2}\left\langle 0\left|m_{u}\bar{u}u+m_{s}\bar{s}s\right|0\right\rangle }{M^{4}}\nonumber \\
 & +\frac{1}{3}\pi^{2}\frac{\left\langle 0\left|\frac{\alpha_{s}}{\pi}G_{\mu\nu}^{a}G_{\mu\nu}^{a}\right|0\right\rangle }{M^{4}}-2\pi^{3}\frac{\left\langle 0\left|\alpha_{s}\left(\bar{u}\gamma_{\alpha}\gamma_{5}t^{a}u-\bar{s}\gamma_{\alpha}\gamma_{5}t^{a}s\right)^{2}\right|0\right\rangle }{M^{6}}\\
 & -\frac{4}{9}\pi^{3}\frac{\left\langle 0\left|\alpha_{s}\left(\bar{u}\gamma_{\alpha}t^{a}u+\bar{s}\gamma_{\alpha}t^{a}s\underset{q=u,d,s}{\sum}\bar{q}\gamma_{\alpha}t^{a}q\right)\right|0\right\rangle }{M^{6}}\biggr]\nonumber
\end{eqnarray}
For the $\varphi$ meson
\begin{eqnarray}
 & \int\mathrm{e}^{-\nicefrac{s}{M^{2}}}\rho^{\left(\varphi\right)}\left(s\right)\mathrm{d}s=\frac{M^{2}}{36\pi}\biggl[1+\frac{\alpha_{s}\left(M\right)}{\pi}-\frac{6m_{s}^{2}\left(M\right)}{M^{2}}+\frac{8\pi^{2}\left\langle 0\left|m_{s}\bar{s}s\right|0\right\rangle }{M^{4}}\nonumber \\
 & +\frac{1}{3}\pi^{2}\frac{\left\langle 0\left|\frac{\alpha_{s}}{\pi}G_{\mu\nu}^{a}G_{\mu\nu}^{a}\right|0\right\rangle }{M^{4}}-\frac{448}{81}\pi^{3}\alpha_{s}\left(\mu\right)\frac{\left\langle 0\left|\bar{q}q\right|0\right\rangle ^{2}}{M^{6}}\biggr]
\end{eqnarray}
For the nucleon \cite{Drukarev2009}
\begin{eqnarray}
\tilde{A}_{0}+\tilde{A}_{4}+\tilde{A}_{6}+\tilde{A}_{8} & = &
\beta_{N}^{2}\mathrm{e}^{-\nicefrac{m^{2}}{M^{2}}}+\beta_{N^{\prime}}^{2}\mathrm{e}^{-\nicefrac{m^{\prime2}}{M^{2}}}
\end{eqnarray}
where
\begin{eqnarray*}
\tilde{A}_{0}\left(M^{2},W^{2}\right) & = & M^{6}E_{2}\left[1+\frac{\alpha_{s}}{\pi}\left(\frac{53}{13}-\mathrm{ln}\frac{W^{2}}{\mu^{2}}\right)\right]\\
 &  & -\frac{\alpha_{s}}{\pi}\left[M^{4}W^{2}\left(1+\frac{3W^{2}}{4M^{2}}\right)\mathrm{e}^{-\frac{W^{2}}{M^{2}}}+M^{6}\varepsilon\left(-\frac{W^{2}}{M^{2}}\right)\right]\\
\tilde{A}_{4}\left(M^{2},W^{2}\right) & = & \frac{bM^{2}E_{0}}{4L}\\
\tilde{A}_{6}\left(M^{2},W^{2}\right) & = &
\frac{4}{3}a^{2}\left[1-\frac{\alpha_{s}}{\pi}\left(\frac{5}{6}+\frac{1}{3}\left(\mathrm{ln}\frac{W^{2}}{\mu^{2}}+\varepsilon\left(-\frac{W^{2}}{M^{2}}\right)\right)\right)\right]
\end{eqnarray*}

\[
a=-\left(2\pi\right)^{2}\left\langle \bar{q}q\right\rangle
,b=\left(2\pi\right)^{2}\left\langle
\frac{\alpha_{s}}{\pi}G^{2}\right\rangle
,\beta_{N}=\left(2\pi\right)^{4}\lambda_{N}^{2},\alpha_{s}\left(1\text{GeV}\right)\approx0.37
\]

\[
E_{0}=1-\mathrm{e}^{-x},E_{2}=1-\left(1+x+\frac{1}{2}x^{2}\right)\mathrm{e}^{-x}
\]
with $x=\nicefrac{W^{2}}{M^{2}}$,
$\varepsilon\left(x\right)=\underset{n}{\sum}\frac{x^{n}}{n\cdot
n!}$,

\[
L=\frac{\ln\left(\nicefrac{M^{2}}{\Lambda^{2}}\right)}{\ln\left(\nicefrac{\mu^{2}}{\Lambda^{2}}\right)}
\]

\bibliographystyle{aipnum4-1}
\bibliography{refs}

\begin{thebibliography}{21}%
\makeatletter
\providecommand \@ifxundefined [1]{%
 \@ifx{#1\undefined}
}%
\providecommand \@ifnum [1]{%
 \ifnum #1\expandafter \@firstoftwo
 \else \expandafter \@secondoftwo
 \fi
}%
\providecommand \@ifx [1]{%
 \ifx #1\expandafter \@firstoftwo
 \else \expandafter \@secondoftwo
 \fi
}%
\providecommand \natexlab [1]{#1}%
\providecommand \enquote  [1]{``#1''}%
\providecommand \bibnamefont  [1]{#1}%
\providecommand \bibfnamefont [1]{#1}%
\providecommand \citenamefont [1]{#1}%
\providecommand \href@noop [0]{\@secondoftwo}%
\providecommand \href [0]{\begingroup \@sanitize@url \@href}%
\providecommand \@href[1]{\@@startlink{#1}\@@href}%
\providecommand \@@href[1]{\endgroup#1\@@endlink}%
\providecommand \@sanitize@url [0]{\catcode `\\12\catcode `\$12\catcode
  `\&12\catcode `\#12\catcode `\^12\catcode `\_12\catcode `\%12\relax}%
\providecommand \@@startlink[1]{}%
\providecommand \@@endlink[0]{}%
\providecommand \url  [0]{\begingroup\@sanitize@url \@url }%
\providecommand \@url [1]{\endgroup\@href {#1}{\urlprefix }}%
\providecommand \urlprefix  [0]{URL }%
\providecommand \Eprint [0]{\href }%
\providecommand \doibase [0]{http://dx.doi.org/}%
\providecommand \selectlanguage [0]{\@gobble}%
\providecommand \bibinfo  [0]{\@secondoftwo}%
\providecommand \bibfield  [0]{\@secondoftwo}%
\providecommand \translation [1]{[#1]}%
\providecommand \BibitemOpen [0]{}%
\providecommand \bibitemStop [0]{}%
\providecommand \bibitemNoStop [0]{.\EOS\space}%
\providecommand \EOS [0]{\spacefactor3000\relax}%
\providecommand \BibitemShut  [1]{\csname bibitem#1\endcsname}%
\let\auto@bib@innerbib\@empty
\bibitem [{\citenamefont {Shifman}, \citenamefont {Vainshtein},\ and\
  \citenamefont {Zakharov}(1979)}]{Shifman1979}%
  \BibitemOpen
  \bibfield  {author} {\bibinfo {author} {\bibfnamefont {M.~A.}\ \bibnamefont
  {Shifman}}, \bibinfo {author} {\bibfnamefont {A.~I.}\ \bibnamefont
  {Vainshtein}}, \ and\ \bibinfo {author} {\bibfnamefont {V.~I.}\ \bibnamefont
  {Zakharov}},\ }\href {\doibase 10.1016/0550-3213(79)90022-1} {\bibfield
  {journal} {\bibinfo  {journal} {Nucl. Phys.}\ }\textbf {\bibinfo {volume}
  {B147}},\ \bibinfo {pages} {385} (\bibinfo {year} {1979})}\BibitemShut
  {NoStop}%
\bibitem [{\citenamefont {Krasnikov}\ and\ \citenamefont
  {Pivovarov}(1982)}]{Krasnikov1982}%
  \BibitemOpen
  \bibfield  {author} {\bibinfo {author} {\bibfnamefont {N.~V.}\ \bibnamefont
  {Krasnikov}}\ and\ \bibinfo {author} {\bibfnamefont {A.~A.}\ \bibnamefont
  {Pivovarov}},\ }\href {\doibase 10.1016/0370-2693(82)91077-2} {\bibfield
  {journal} {\bibinfo  {journal} {Phys. Lett.}\ }\textbf {\bibinfo {volume}
  {B112}},\ \bibinfo {pages} {397} (\bibinfo {year} {1982})},\ \bibinfo {note}
  {[Yad. Fiz.35,1270(1982)]}\BibitemShut {NoStop}%
\bibitem [{\citenamefont {Kataev}, \citenamefont {Krasnikov},\ and\
  \citenamefont {Pivovarov}(1983)}]{Kataev1983}%
  \BibitemOpen
  \bibfield  {author} {\bibinfo {author} {\bibfnamefont {A.~L.}\ \bibnamefont
  {Kataev}}, \bibinfo {author} {\bibfnamefont {N.~V.}\ \bibnamefont
  {Krasnikov}}, \ and\ \bibinfo {author} {\bibfnamefont {A.~A.}\ \bibnamefont
  {Pivovarov}},\ }\bibfield  {booktitle} {\emph {\bibinfo {booktitle} {{Quarks
  82 Sukhumi, USSR, May 5-7, 1982}}},\ }\href {\doibase
  10.1016/0370-2693(83)90966-8} {\bibfield  {journal} {\bibinfo  {journal}
  {Phys. Lett.}\ }\textbf {\bibinfo {volume} {B123}},\ \bibinfo {pages} {93}
  (\bibinfo {year} {1983})}\BibitemShut {NoStop}%
\bibitem [{\citenamefont {Gorishnii}, \citenamefont {Kataev},\ and\
  \citenamefont {Larin}(1984)}]{Gorishnii1984}%
  \BibitemOpen
  \bibfield  {author} {\bibinfo {author} {\bibfnamefont {S.~G.}\ \bibnamefont
  {Gorishnii}}, \bibinfo {author} {\bibfnamefont {A.~L.}\ \bibnamefont
  {Kataev}}, \ and\ \bibinfo {author} {\bibfnamefont {S.~A.}\ \bibnamefont
  {Larin}},\ }\href {\doibase 10.1016/0370-2693(84)90315-0} {\bibfield
  {journal} {\bibinfo  {journal} {Phys. Lett.}\ }\textbf {\bibinfo {volume}
  {B135}},\ \bibinfo {pages} {457} (\bibinfo {year} {1984})}\BibitemShut
  {NoStop}%
\bibitem [{\citenamefont {Gelhausen}\ \emph {et~al.}(2014)\citenamefont
  {Gelhausen}, \citenamefont {Khodjamirian}, \citenamefont {Pivovarov},\ and\
  \citenamefont {Rosenthal}}]{Gelhausen2014}%
  \BibitemOpen
  \bibfield  {author} {\bibinfo {author} {\bibfnamefont {P.}~\bibnamefont
  {Gelhausen}}, \bibinfo {author} {\bibfnamefont {A.}~\bibnamefont
  {Khodjamirian}}, \bibinfo {author} {\bibfnamefont {A.~A.}\ \bibnamefont
  {Pivovarov}}, \ and\ \bibinfo {author} {\bibfnamefont {D.}~\bibnamefont
  {Rosenthal}},\ }\href {\doibase 10.1140/epjc/s10052-014-2979-z} {\bibfield
  {journal} {\bibinfo  {journal} {Eur. Phys. J.}\ }\textbf {\bibinfo {volume}
  {C74}},\ \bibinfo {pages} {2979} (\bibinfo {year} {2014})},\ \Eprint
  {http://arxiv.org/abs/1404.5891} {arXiv:1404.5891 [hep-ph]} \BibitemShut
  {NoStop}%
\bibitem [{\citenamefont {Olive}\ \emph {et~al.}(2014)\citenamefont {Olive}
  \emph {et~al.}}]{Olive2014}%
  \BibitemOpen
  \bibfield  {author} {\bibinfo {author} {\bibfnamefont {K.~A.}\ \bibnamefont
  {Olive}} \emph {et~al.} (\bibinfo {collaboration} {Particle Data Group}),\
  }\href {\doibase 10.1088/1674-1137/38/9/090001} {\bibfield  {journal}
  {\bibinfo  {journal} {Chin. Phys.}\ }\textbf {\bibinfo {volume} {C38}},\
  \bibinfo {pages} {090001} (\bibinfo {year} {2014})}\BibitemShut {NoStop}%
\bibitem [{\citenamefont {Narison}(1984)}]{Narison1984}%
  \BibitemOpen
  \bibfield  {author} {\bibinfo {author} {\bibfnamefont {S.}~\bibnamefont
  {Narison}},\ }\href {\doibase 10.1007/BF01421756} {\bibfield  {journal}
  {\bibinfo  {journal} {Z. Phys.}\ }\textbf {\bibinfo {volume} {C26}},\
  \bibinfo {pages} {209} (\bibinfo {year} {1984})}\BibitemShut {NoStop}%
\bibitem [{\citenamefont {Drukarev}, \citenamefont {Ryskin},\ and\
  \citenamefont {Sadovnikova}(2009)}]{Drukarev2009}%
  \BibitemOpen
  \bibfield  {author} {\bibinfo {author} {\bibfnamefont {E.~G.}\ \bibnamefont
  {Drukarev}}, \bibinfo {author} {\bibfnamefont {M.~G.}\ \bibnamefont
  {Ryskin}}, \ and\ \bibinfo {author} {\bibfnamefont {V.~A.}\ \bibnamefont
  {Sadovnikova}},\ }\href {\doibase 10.1103/PhysRevD.80.014008} {\bibfield
  {journal} {\bibinfo  {journal} {Phys. Rev.}\ }\textbf {\bibinfo {volume}
  {D80}},\ \bibinfo {pages} {014008} (\bibinfo {year} {2009})},\ \Eprint
  {http://arxiv.org/abs/0812.4347} {arXiv:0812.4347 [nucl-th]} \BibitemShut
  {NoStop}%
\bibitem [{\citenamefont {Ioffe}(2003)}]{Ioffe2003}%
  \BibitemOpen
  \bibfield  {author} {\bibinfo {author} {\bibfnamefont {B.~L.}\ \bibnamefont
  {Ioffe}},\ }\href {\doibase 10.1134/1.1540654} {\bibfield  {journal}
  {\bibinfo  {journal} {Phys. Atom. Nucl.}\ }\textbf {\bibinfo {volume} {66}},\
  \bibinfo {pages} {30} (\bibinfo {year} {2003})},\ \bibinfo {note} {[Yad.
  Fiz.66,32(2003)]},\ \Eprint {http://arxiv.org/abs/hep-ph/0207191}
  {arXiv:hep-ph/0207191 [hep-ph]} \BibitemShut {NoStop}%
\bibitem [{\citenamefont {Becirevic}\ \emph {et~al.}(2003)\citenamefont
  {Becirevic}, \citenamefont {Lubicz}, \citenamefont {Mescia},\ and\
  \citenamefont {Tarantino}}]{Becirevic2003}%
  \BibitemOpen
  \bibfield  {author} {\bibinfo {author} {\bibfnamefont {D.}~\bibnamefont
  {Becirevic}}, \bibinfo {author} {\bibfnamefont {V.}~\bibnamefont {Lubicz}},
  \bibinfo {author} {\bibfnamefont {F.}~\bibnamefont {Mescia}}, \ and\ \bibinfo
  {author} {\bibfnamefont {C.}~\bibnamefont {Tarantino}},\ }\href {\doibase
  10.1088/1126-6708/2003/05/007} {\bibfield  {journal} {\bibinfo  {journal}
  {JHEP}\ }\textbf {\bibinfo {volume} {05}},\ \bibinfo {pages} {007} (\bibinfo
  {year} {2003})},\ \Eprint {http://arxiv.org/abs/hep-lat/0301020}
  {arXiv:hep-lat/0301020 [hep-lat]} \BibitemShut {NoStop}%
\bibitem [{\citenamefont {Wingate}\ \emph {et~al.}(1995)\citenamefont
  {Wingate}, \citenamefont {DeGrand}, \citenamefont {Collins},\ and\
  \citenamefont {Heller}}]{Wingate1995}%
  \BibitemOpen
  \bibfield  {author} {\bibinfo {author} {\bibfnamefont {M.}~\bibnamefont
  {Wingate}}, \bibinfo {author} {\bibfnamefont {T.}~\bibnamefont {DeGrand}},
  \bibinfo {author} {\bibfnamefont {S.}~\bibnamefont {Collins}}, \ and\
  \bibinfo {author} {\bibfnamefont {U.~M.}\ \bibnamefont {Heller}},\ }\href
  {\doibase 10.1103/PhysRevLett.74.4596} {\bibfield  {journal} {\bibinfo
  {journal} {Phys. Rev. Lett.}\ }\textbf {\bibinfo {volume} {74}},\ \bibinfo
  {pages} {4596} (\bibinfo {year} {1995})}\BibitemShut {NoStop}%
\bibitem [{\citenamefont {Andrianov}, \citenamefont {Espriu},\ and\
  \citenamefont {Tarrach}(1998)}]{Andrianov1998}%
  \BibitemOpen
  \bibfield  {author} {\bibinfo {author} {\bibfnamefont {A.~A.}\ \bibnamefont
  {Andrianov}}, \bibinfo {author} {\bibfnamefont {D.}~\bibnamefont {Espriu}}, \
  and\ \bibinfo {author} {\bibfnamefont {R.}~\bibnamefont {Tarrach}},\ }\href
  {\doibase 10.1016/S0550-3213(98)00508-2} {\bibfield  {journal} {\bibinfo
  {journal} {Nucl. Phys.}\ }\textbf {\bibinfo {volume} {B533}},\ \bibinfo
  {pages} {429} (\bibinfo {year} {1998})},\ \Eprint
  {http://arxiv.org/abs/hep-ph/9803232} {arXiv:hep-ph/9803232 [hep-ph]}
  \BibitemShut {NoStop}%
\bibitem [{\citenamefont {Elias}\ \emph {et~al.}(1997)\citenamefont {Elias},
  \citenamefont {Fariborz}, \citenamefont {Samuel}, \citenamefont {Shi},\ and\
  \citenamefont {Steele}}]{Elias1997}%
  \BibitemOpen
  \bibfield  {author} {\bibinfo {author} {\bibfnamefont {V.}~\bibnamefont
  {Elias}}, \bibinfo {author} {\bibfnamefont {A.}~\bibnamefont {Fariborz}},
  \bibinfo {author} {\bibfnamefont {M.~A.}\ \bibnamefont {Samuel}}, \bibinfo
  {author} {\bibfnamefont {F.}~\bibnamefont {Shi}}, \ and\ \bibinfo {author}
  {\bibfnamefont {T.~G.}\ \bibnamefont {Steele}},\ }\href {\doibase
  10.1016/S0370-2693(97)01049-6} {\bibfield  {journal} {\bibinfo  {journal}
  {Phys. Lett.}\ }\textbf {\bibinfo {volume} {B412}},\ \bibinfo {pages} {131}
  (\bibinfo {year} {1997})},\ \Eprint {http://arxiv.org/abs/hep-ph/9706472}
  {arXiv:hep-ph/9706472 [hep-ph]} \BibitemShut {NoStop}%
\bibitem [{\citenamefont {Maltman}\ and\ \citenamefont
  {Kambor}(2002)}]{Maltman2002}%
  \BibitemOpen
  \bibfield  {author} {\bibinfo {author} {\bibfnamefont {K.}~\bibnamefont
  {Maltman}}\ and\ \bibinfo {author} {\bibfnamefont {J.}~\bibnamefont
  {Kambor}},\ }\href {\doibase 10.1103/PhysRevD.65.074013} {\bibfield
  {journal} {\bibinfo  {journal} {Phys. Rev.}\ }\textbf {\bibinfo {volume}
  {D65}},\ \bibinfo {pages} {074013} (\bibinfo {year} {2002})},\ \Eprint
  {http://arxiv.org/abs/hep-ph/0108227} {arXiv:hep-ph/0108227 [hep-ph]}
  \BibitemShut {NoStop}%
\bibitem [{\citenamefont {McNeile}\ and\ \citenamefont
  {Michael}(2006)}]{McNeile2006}%
  \BibitemOpen
  \bibfield  {author} {\bibinfo {author} {\bibfnamefont {C.}~\bibnamefont
  {McNeile}}\ and\ \bibinfo {author} {\bibfnamefont {C.}~\bibnamefont
  {Michael}} (\bibinfo {collaboration} {UKQCD}),\ }\href {\doibase
  10.1016/j.physletb.2006.09.056} {\bibfield  {journal} {\bibinfo  {journal}
  {Phys. Lett.}\ }\textbf {\bibinfo {volume} {B642}},\ \bibinfo {pages} {244}
  (\bibinfo {year} {2006})},\ \Eprint {http://arxiv.org/abs/hep-lat/0607032}
  {arXiv:hep-lat/0607032 [hep-lat]} \BibitemShut {NoStop}%
\bibitem [{\citenamefont {Volkov}\ and\ \citenamefont
  {Weiss}(1997)}]{Volkov1997}%
  \BibitemOpen
  \bibfield  {author} {\bibinfo {author} {\bibfnamefont {M.~K.}\ \bibnamefont
  {Volkov}}\ and\ \bibinfo {author} {\bibfnamefont {C.}~\bibnamefont {Weiss}},\
  }\href {\doibase 10.1103/PhysRevD.56.221} {\bibfield  {journal} {\bibinfo
  {journal} {Phys. Rev.}\ }\textbf {\bibinfo {volume} {D56}},\ \bibinfo {pages}
  {221} (\bibinfo {year} {1997})},\ \Eprint
  {http://arxiv.org/abs/hep-ph/9608347} {arXiv:hep-ph/9608347 [hep-ph]}
  \BibitemShut {NoStop}%
\bibitem [{\citenamefont {Holl}, \citenamefont {Krassnigg},\ and\ \citenamefont
  {Roberts}(2004)}]{Holl2004}%
  \BibitemOpen
  \bibfield  {author} {\bibinfo {author} {\bibfnamefont {A.}~\bibnamefont
  {Holl}}, \bibinfo {author} {\bibfnamefont {A.}~\bibnamefont {Krassnigg}}, \
  and\ \bibinfo {author} {\bibfnamefont {C.~D.}\ \bibnamefont {Roberts}},\
  }\href {\doibase 10.1103/PhysRevC.70.042203} {\bibfield  {journal} {\bibinfo
  {journal} {Phys. Rev.}\ }\textbf {\bibinfo {volume} {C70}},\ \bibinfo {pages}
  {042203} (\bibinfo {year} {2004})},\ \Eprint
  {http://arxiv.org/abs/nucl-th/0406030} {arXiv:nucl-th/0406030 [nucl-th]}
  \BibitemShut {NoStop}%
\bibitem [{\citenamefont {Holl}\ \emph {et~al.}(2005)\citenamefont {Holl},
  \citenamefont {Krassnigg}, \citenamefont {Maris}, \citenamefont {Roberts},\
  and\ \citenamefont {Wright}}]{Holl2005}%
  \BibitemOpen
  \bibfield  {author} {\bibinfo {author} {\bibfnamefont {A.}~\bibnamefont
  {Holl}}, \bibinfo {author} {\bibfnamefont {A.}~\bibnamefont {Krassnigg}},
  \bibinfo {author} {\bibfnamefont {P.}~\bibnamefont {Maris}}, \bibinfo
  {author} {\bibfnamefont {C.~D.}\ \bibnamefont {Roberts}}, \ and\ \bibinfo
  {author} {\bibfnamefont {S.~V.}\ \bibnamefont {Wright}},\ }\href {\doibase
  10.1103/PhysRevC.71.065204} {\bibfield  {journal} {\bibinfo  {journal} {Phys.
  Rev.}\ }\textbf {\bibinfo {volume} {C71}},\ \bibinfo {pages} {065204}
  (\bibinfo {year} {2005})},\ \Eprint {http://arxiv.org/abs/nucl-th/0503043}
  {arXiv:nucl-th/0503043 [nucl-th]} \BibitemShut {NoStop}%
\bibitem [{\citenamefont {Qin}\ \emph {et~al.}(2012)\citenamefont {Qin},
  \citenamefont {Chang}, \citenamefont {Liu}, \citenamefont {Roberts},\ and\
  \citenamefont {Wilson}}]{Qin2012}%
  \BibitemOpen
  \bibfield  {author} {\bibinfo {author} {\bibfnamefont {S.-x.}\ \bibnamefont
  {Qin}}, \bibinfo {author} {\bibfnamefont {L.}~\bibnamefont {Chang}}, \bibinfo
  {author} {\bibfnamefont {Y.-x.}\ \bibnamefont {Liu}}, \bibinfo {author}
  {\bibfnamefont {C.~D.}\ \bibnamefont {Roberts}}, \ and\ \bibinfo {author}
  {\bibfnamefont {D.~J.}\ \bibnamefont {Wilson}},\ }\href {\doibase
  10.1103/PhysRevC.85.035202} {\bibfield  {journal} {\bibinfo  {journal} {Phys.
  Rev.}\ }\textbf {\bibinfo {volume} {C85}},\ \bibinfo {pages} {035202}
  (\bibinfo {year} {2012})},\ \Eprint {http://arxiv.org/abs/1109.3459}
  {arXiv:1109.3459 [nucl-th]} \BibitemShut {NoStop}%
\bibitem [{\citenamefont {Narison}(2014)}]{Narison2014}%
  \BibitemOpen
  \bibfield  {author} {\bibinfo {author} {\bibfnamefont {S.}~\bibnamefont
  {Narison}},\ }\href {\doibase 10.1016/j.physletb.2014.09.056} {\bibfield
  {journal} {\bibinfo  {journal} {Phys. Lett.}\ }\textbf {\bibinfo {volume}
  {B738}},\ \bibinfo {pages} {346} (\bibinfo {year} {2014})},\ \Eprint
  {http://arxiv.org/abs/1401.3689} {arXiv:1401.3689 [hep-ph]} \BibitemShut
  {NoStop}%
\bibitem [{\citenamefont {Dominguez}(1977)}]{Dominguez1977}%
  \BibitemOpen
  \bibfield  {author} {\bibinfo {author} {\bibfnamefont {C.~A.}\ \bibnamefont
  {Dominguez}},\ }\href {\doibase 10.1103/PhysRevD.15.1350} {\bibfield
  {journal} {\bibinfo  {journal} {Phys. Rev.}\ }\textbf {\bibinfo {volume}
  {D15}},\ \bibinfo {pages} {1350} (\bibinfo {year} {1977})}\BibitemShut
  {NoStop}%
\end{thebibliography}%

\end{document}